\documentclass[runningheads, envcountsame, a4paper]{llncs}
\usepackage{siunitx}
\usepackage{float}
\usepackage{amsmath}
\usepackage{soul}
\usepackage{color}
\usepackage[pdftex]{graphicx}
\usepackage{epstopdf}
\usepackage[hyphens]{url}
\usepackage[misc]{ifsym}

\usepackage{tabularx, ragged2e}
\newcolumntype{C}{>{\Centering\arraybackslash}X}
\begin{document}

\title{Deep Residual Local Feature Learning for Speech Emotion Recognition}

\titlerunning{DeepResLFLB for Speech Emotion Recognition}
%

\toctitle{Deep Residual Local Feature Learning for Speech Emotion Recognition}
\tocauthor{Sattaya Singkul~Thakorn Chatchaisathaporn~Boontawee Suntisrivaraporn~Kuntpong Woraratpanya\  }

\author{Sattaya Singkul\inst{1}\orcidID{0000-0001-7335-7105} \and
Thakorn Chatchaisathaporn\inst{2}
\newline Boontawee Suntisrivaraporn\inst{2}\and Kuntpong Woraratpanya\inst{1,\thanks{Corresponding author}}\ }
\authorrunning{S. Singkul et al.}
\institute{Faculty of Information Technology, \newline King Mongkut's Institute of Technology Ladkrabang, Bangkok, Thailand\\
\email{\{59070173,kuntpong\}@it.kmitl.ac.th}
 \and
Data Analytics, Siam Commercial Bank, Bangkok, Thailand\\
\email{thakorn.chatchaisathaporn@scb.co.th},\
 \email{meng234@gmail.com}}


\maketitle              
\begin{abstract}
Speech Emotion Recognition (SER) is becoming a key role in global business today to improve service efficiency, like call center services. Recent SERs were based on a deep learning approach. However, the efficiency of deep learning depends on the number of layers, i.e., the deeper layers, the higher efficiency. On the other hand, the deeper layers are causes of a vanishing gradient problem, a low learning rate, and high time-consuming. Therefore, this paper proposed a redesign of existing local feature learning block (LFLB). The new design is called a deep residual local feature learning block (DeepResLFLB). DeepResLFLB consists of three cascade blocks: LFLB, residual local feature learning block (ResLFLB), and multilayer perceptron (MLP). LFLB is built for learning local correlations along with extracting hierarchical correlations; DeepResLFLB can take advantage of repeatedly learning to explain more detail in deeper layers using residual learning for solving vanishing gradient and reducing overfitting; and MLP is adopted to find the relationship of learning and discover probability for predicted speech emotions and gender types. Based on two available published datasets: EMODB·and RAVDESS, the proposed DeepResLFLB can significantly improve performance when evaluated by standard metrics: accuracy, precision, recall, and F1-score.

\end{abstract}

\section{Introduction}
Emotional analysis has been an active research area for a few decades, especially in recognition domains of text and speech emotions. Even if text and speech emotions are closely relevant, both kinds of emotions have different challenges. One of the challenges in text emotion recognition is ambiguous words, resulting from omitted words \cite{9045639,8930002}. On the other hand, one of the challenges in speech emotion recognition is creating an efficient model. However, this paper focuses on only the recognition of speech emotions. In this area, two types of information, linguistic and paralinguistic, were mainly considered in speech emotion recognition. The linguistic information refers to the meaning or context of speech. The paralinguistic information implies the implicit message meaning, like the emotion in speech \cite{el2011survey,anagnostopoulos2015features,zhang2014cooperative,guidi2015automatic}. Speech characteristics can interpret the meaning of speech; therefore, behavioral expression was investigated in most of the speech emotion recognition works  \cite{gunes2007bi,bong2017implementation,yuvaraj2014detection}.
\vspace{-1pt}

In recent works, local feature learning block (LFLB) \cite{zhao2019speech}, one of the efficient methods, has been used in integrating local and global speech emotion features, which provide better results in recognition. Inside LFLB, convolution neural network (CNN) was used for extracting local features, and then long short-term memory (LSTM) was applied for extracting contextual dependencies from those local features to learn in a time-related relationship. However, vanishing gradient problems may occur with CNN \cite{he2016deep}. Therefore, residual deep learning was applied to the CNN by using skip-connection to reduce unnecessary learning and add feature details that may be lost in between layers.

Furthermore, the accuracy of speech recognition does not only rely on the efficiency of a model, but also of a speech feature selection \cite{wu2011automatic}. In terms of speech characteristics, there are many distinctive acoustic features that usually used in recognizing the speech emotion, such as continuous features, qualitative features, and spectral features \cite{he2011study,wu2011automatic,perez2012acoustic,huang2014speech,huang2015extraction}. Many of them have been investigated to recognize speech emotions. Some researchers compared the pros and cons of each feature, but no one can identify which feature was the best one until now \cite{el2011survey,anagnostopoulos2015features,demircan2018application,sun2015weighted}.

As previously mentioned, we proposed a method to improve the efficiency of LFLB \cite{he2016deep} for deeper learning. The proposed method, deep residual local feature learning block (DeepResLFLB), was inspired by the concept of human brain learning; that is, ‘repeated reading makes learning more effective,’ as the same way that Sari \cite{sari2019influence} and Shanahan \cite{shanahan2017repeatedread} were used. Responding to our inspired concept, we implemented a learning method for speech emotion recognition with three parts: Part 1 is for general learning, like human reading for the first time, Part 2 is for further learning, like additional readings, and the last part is for associating parts learned to decide types of emotions. Besides, the feature selection is compared with two types of distinctive features to find the most effective feature in our work: the normal and specific distinctive features are log-mel spectrogram (LMS), which is fully filtered sound elements, and 
MFCC deltas, delta-deltas, and chromagram (LMSDDC) are more clearly identify speech characteristics extracted based on 
the human mood.

Our main contributions of this paper are as follows: 
(i) Deep residual local feature learning block (DeepResLFLB) was proposed. DeepResLFLB was arranged its internal network as LFLB, batch normalization (BN), activation function, normalization-activation-CNN (NAC), and deep layers. 
(ii) Learning sequences of DeepResLFLB were imitated from human re-reads. 
(iii) Speech emotion features, 
based on human mood determination factors such as LMS and LMSDDC, were applied and compared their performances. 

\section{Literature Reviews}
Model efficiency is one of the important factors in SER. Many papers focused on learning methods of machine learning or deep learning. Demircan \cite{demircan2018application} introduced fuzzy c-mean as a preprocessing step to group and add characteristics before using machine learning. 
Venkataramanan \cite{venkataramanan2019emotion} studied of using deep learning in SER. The findings of the study revealed that CNN outperformed the traditional machine learning. Also, Huang \cite{huang2014speech} showed that semi-CNN in SER can increase accuracy. Zhao \cite{zhao2019speech} presented the use of CNN in conjunction with LSTM to extract and learn features. Zhao’s method used a sequence of CNN in a block style, consisting of CNN, BN, activation function, and pooling, for local feature learning, and then used LSTM for extracting contextual dependencies in a time-related relationship. In this way, both local and global features are extracted and learned.   

It is undeniable that the effectiveness of deep learning mainly depends on the data size for training \cite{impactdatasize}. Recently, Google brain research \cite{park2019specaugment} proposed data augmentation, one of the efficient techniques that can increase the amount of data, by adding spectrogram characteristics, also known as “Spectrogram Augmentation.” This augmentation consists of time warping to see more time shift patterns, time masking to reduce the overfitting rate of the model and improve the sound tolerance that may have characteristics of silence, frequency masking to reduce the overfitting rate and increase sound resistance from concealing characteristics of a specific wavelength. The spectrogram is a basic feature of sound that can lead to various specific features. Therefore, by using above methods, the model can learn more perspectives of the data.

Also, different features lead to different performances in speech emotion recognition. Among the features of speech, mel-frequency cepstral coefficient (MFCC) \cite{venkataramanan2019emotion}, which can be characterized by the frequency filter in the range of 20 Hz to 20 kHz, similar to human hearing, is widely used to obtain coefficients from the filtered sound. Recent research papers \cite{venkataramanan2019emotion,demircan2018application} used the difference of MFCC to get more specific details, but, in the aspect of MFCC, it has no time relationship. Therefore, many papers \cite{venkataramanan2019emotion,zhao2019speech,huang2014speech} used mel spectrogram (MS) instead. MS can respond to the time relationship, thus providing better results than just using MFCC. Besides, music can be looked different from speech; therefore, chromagram is widely used instead of MFCC, since it can provide better features than normal MFCC and MS.

Our work is different from the previously mentioned works in that the deep residual local feature learning block (DeepResLFLB) was redesigned from LFLB. 
This method helps reduce the chance of feature and updated losses caused by CNN model in the LFLB, especially in deeper layers. DeepResLFLB uses a repeated learning style that local features extracted from a bias frame with silent voice (see subsection 3.3) can be learned through a residual deep learning approach. Moreover, we extracted distinctive features based on a concept of determining human emotions, consisting of prosodic \cite{8250527}, filter bank \cite{8250527,venkataramanan2019emotion}, and glottal flow \cite{degottex2010glottal}. These three features in conjunction with our ResLFLB can improve learning efficiency.

\section{The Proposed Model}
\begin{figure}
\vspace{-20pt}
\includegraphics[width=\textwidth]{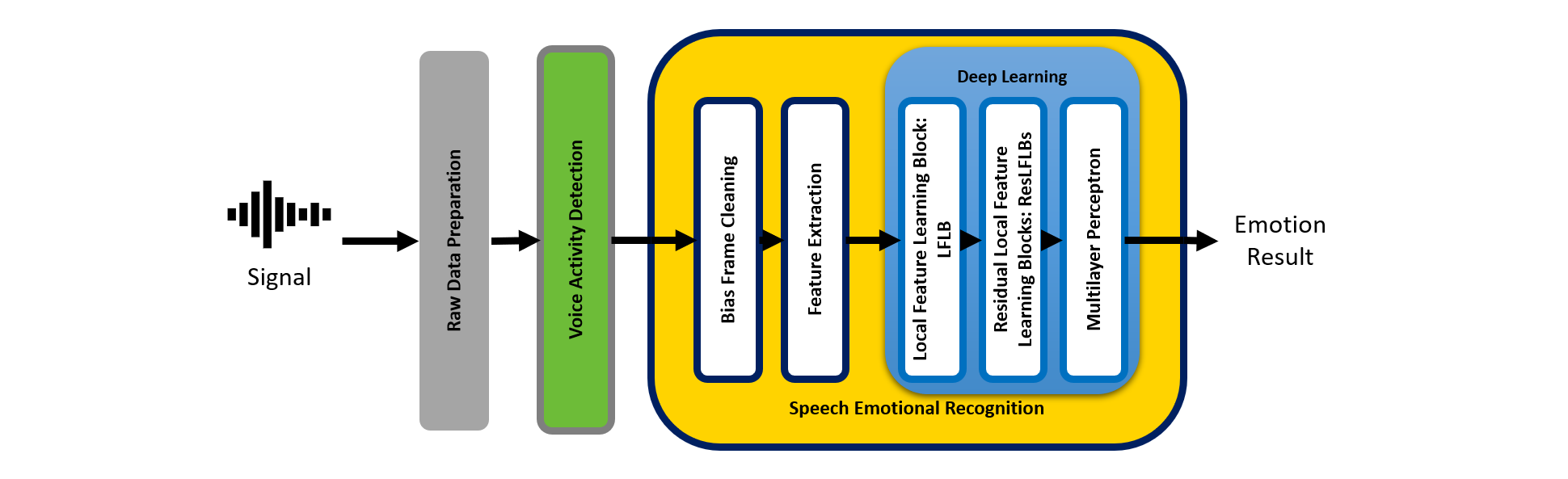}
\caption{A deep residual local feature learning framework.} \label{framwork}
\vspace{-10pt}
\end{figure}

To enable SER as efficiently as possible, the following factors: raw datasets, environments, and features are included in our system design. Based on such factors, a new designed framework, called DeepResLFLB, was proposed as shown in Fig.~\ref{framwork}. This framework consists of five parts: (i) raw data preparation, (ii) voice activity detection, (iii) bias frame cleaning, (iv) feature extraction, and (v) deep learning. 

\subsection{Raw Data Preparation}
Due to the complex nature of datasets and the difference of languages like EMODB in Berlin German and RAVDESS in English, the representation of 
the original datasets may not be enough for training a model based on deep learning. Therefore, increasing a variety of data to see more new dimensions or characteristics is essential. Responding to this, various data augmentation techniques, including noise adding, pitch tuning, and spectrogram, were used in this work to make the model more robust to noise and unseen voice patterns.

\subsection{Voice Activity Detection}
Although both datasets, EMODB and RAVDESS, were produced in a closed environment, quite a bit of noise, they were found that noise remains at the starting and stopping points of sound records. Indeed, noise is not related to speakers’ voice, so it could be removed. Here, voice activity detection \cite{doukhan2018open} was used to detect only voice locations, i.e., excluding noise locations. As a result of the voice activity detection, selected frames can be efficiently analyzed and classified male and female voices by energy-base features. 

\subsection{Bias Frame Cleaning}
Bias frame cleaning is used as a postprocessing of voice activity detection; that is, each frame segmented by the voice activity detection is identified its loudness through Fourier transform (FT). If FT coefficients of a segmented frame are zero, that frame is identified as no significant information for emotional analysis, so it is rejected. 

\subsection{Feature Extraction}
\begin{figure}
\vspace{-10pt}
\includegraphics[width=\textwidth]{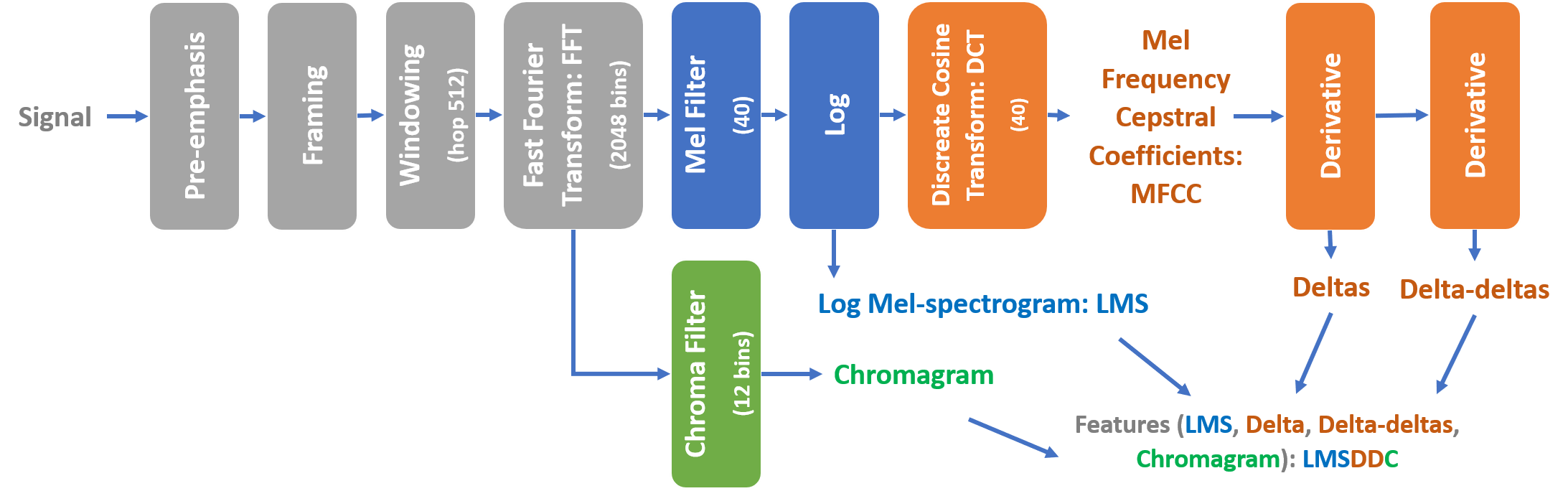}
\caption{Feature extraction of LMS and LMSDDC} \label{feature}
\vspace{-15pt}
\end{figure}

Model performance of deep learning mainly depends on features. The good features usually gain more model performance. Thus, this paper focuses on efficient extraction of human emotion features. Naturally, speech signals always contain human emotions. In other words, we can extract human emotions from speech signals. Here, we briefly describe three important components of speech signals: glottal flow, prosody, and human hearing. Glottal flow can be viewed as a source of speech signals \cite{degottex2010glottal}. It mainly produces fundamental frequencies \cite{doval2006spectrum} or latent sounds within the speech. Prosody is vocal frequencies, which are produced from the air pushed by the lung \cite{degottex2010glottal}. It contains important characteristics, such as intonation, tone, stress, and rhythm. On the other hand, for human hearing, MFCC is one of the analytical tools that can mimic the behavior of human ears by applying cepstral analysis \cite{wang2008recognizing}. Based on our assumption of extracting better emotion features, two important factors are included for feature extraction design: (i) the wide band frequencies of speech signals are regarded as much as possible to cover important features of speech emotions, and (ii) time-frequency processing is used for extracting speech emotions. Here, log-mel spectrogram (LMS) was used as time-frequency representation for emotion features. Two additional features extracted based on MFCC were delta and delta-delta. Furthermore, chromagram feature \cite{robinson1995stimulus,wakefield1999chromagram} was extracted as one of the emotion features. Fig.~\ref{feature} shows our emotion feature extraction. As a result, four features, LMS, delta, delta-delta, and chromagram were used as emotional representation.

\subsection{Deep Learning}

\begin{figure}
\vspace{-10pt}
\includegraphics[width=\textwidth]{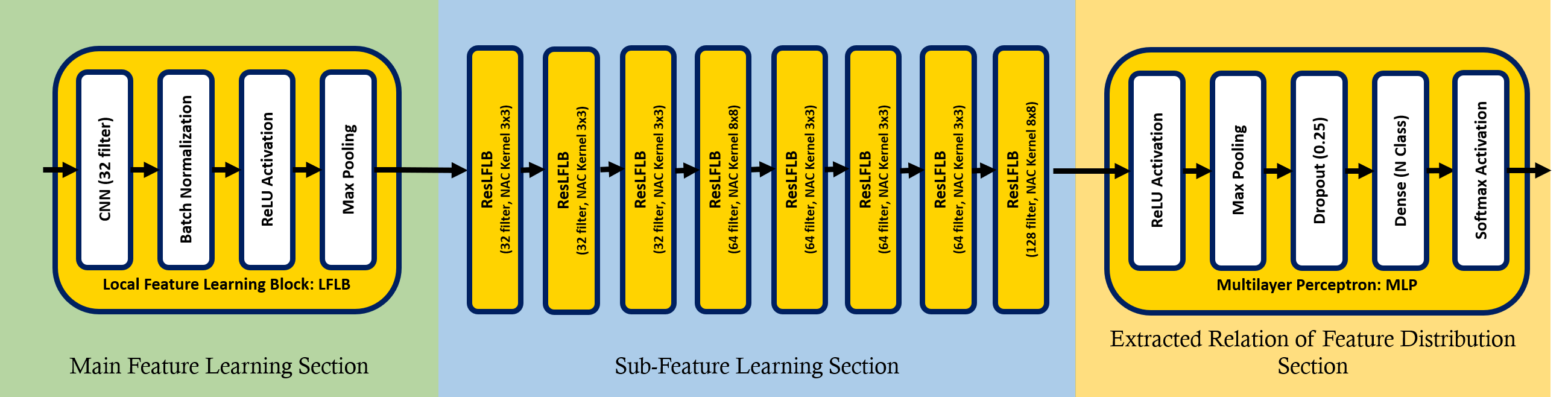}
\caption{A feature learning structure based on the re-reading theory.} \label{reslflb}
\vspace{-10pt}
\end{figure}

Inspired by learning characteristics of human brain activity, i.e., the more repeated reading, the more comprehension. It is similar with Shanahan’s definition, called “repeatedly reads” or “re-reads” \cite{shanahan2017repeatedread}. Responding to the use of re-reading theory for improving the accuracy of SER, we designed a feature learning method as shown in Fig.~\ref{reslflb}, consisting of three sections: (i) main feature learning (MFL), (ii) sub-feature learning (SFL), and (iii) extracted relation of feature distribution (ERFD). 

\subsubsection{Main Feature Learning (MFL) Section}
was designed based on behavior of human brain, like first reading that a human brain starts to learn things. We designed MFLS similar with the LFLB procedure to learn locally basic information as the following steps: (i) 2D-CNN was used to extract necessary features; (ii) BN was applied to enhance learning efficiency of a model; (iii) activation functions converted data to suit for the learning model; and (iv) pooling was for reducing feature size and increased learning speed.

\subsubsection{Sub-Feature Learning (SFL) Section} 
was a further learning process that plays a role in assembling repeated reading for deeper learning. In general, LFLB may be at risk of a vanishing gradient problem that affects learning efficiency. Therefore, we have improved the LFLB’s efficient by means of residual deep learning, or also known as skipping connections, to skip deeper learning layers that are unnecessary and add more feature details after passing each learning layer; this can avoid the vanishing gradient problem. 

\begin{figure}[hbt!]
\vspace{-10pt}
\includegraphics[width=\textwidth]{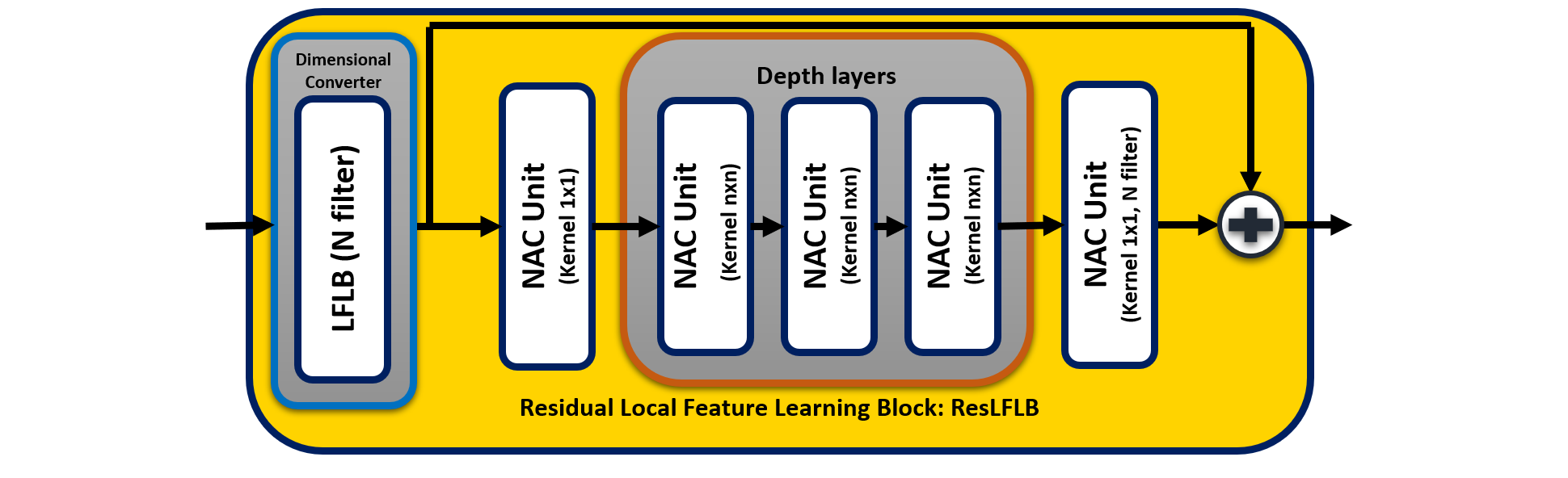}
\caption{A residual local feature learning block structure.} \label{reslflb1}
\vspace{-10pt}
\end{figure}
For sub-feature learning, there are two important sections: (i) A normal LFLB was used as a preprocessing phase of block to transform input data dimensions into a suitable form for the next section. (ii) Depth layers learned features in more depth by using the network sequence as normalization-activation-CNN (NAC) \cite{he2016identity}, which is a sequence of sub-layers in residual deep learning. With this structure, it can reduce test error. In addition, more deeper learning layers may be at risk of a vanishing gradient problem; therefore, the skipping connection was added at an input of this section to compensate features lost in deeper layers. This arrangement of two sections is called residual local feature block (ResLFLB) as shown in Fig.~\ref{reslflb1}. The LFLB also known as a dimensional converter and the NAC final layer had the same output filters to determine the number of output filters of ResLFLB. The kernel size in the first and last layer in deep layers were one. Furthermore, in between the first and last learning layers, the bottleneck design, the compression/decompression strategy, was applied to achieve higher learning performance.
\subsubsection{Extracted Relation of Feature Distribution (ERFD) Section}
was implemented with multilayer perceptron (MLP) to extract the relationship of learning results. A sequence of processes are as follows: (i) Activation ReLU transformed the data to the suitable relationship; (ii) Max pooling reduced data size and extracted important data based on maximum values; (iii) Dropout reduced the overfitting 
of the model; (iv) Dense layers obtained relationships in which neurons equals the number of classes needed to predict; and (v) Activation softmax determined the probability of predicting the emotions.

\section{Experiments and Discussion}
The proposed DeepResLFLB and LMSDDC were evaluated with two main objectives: (i) classification performance and (ii) model performance. In classification performance, four metrics, accuracy, precision, recall, and F1-score as defined by \eqref{acc_eq}, \eqref{pre_eq}, \eqref{re_eq}, and \eqref{f1_eq}, respectively, were used for evaluation. In model performance, validation loss was used as monitoring vanishing gradient problems and the number of CNN layer parameters was used as indicating resource-consuming. The experiments were conducted in comparison with three models: the normal ML with fuzzy c-mean \cite{demircan2018application}, traditional LFLB \cite{zhao2019speech}, and DeepResLFLB, and two different features: LMS and LMSDDC. Note that, in Tables 2 and 4, Dermircan's method was excluded from those experiments due to a mismatch of feature dimensions.

\begin{equation}\label{acc_eq}
Accuracy\:=\:\frac{true\:positive\:+\:true\:negative}{number\:of\:data}
\end{equation}
\begin{equation}\label{pre_eq}
Precision\:=\:\frac{true\:positive}{total\:predicted\:positive}
\end{equation}
\begin{equation}\label{re_eq}
Recall\:=\:\frac{true\:positive}{total\:actual\:positive}
\end{equation}
\begin{equation}\label{f1_eq}
F1\:=\:2\times\frac{precision\:\times\:recall}{precision\:+\:recall}
\end{equation}

\subsubsection{Dataset preparation}
Two available published datasets: Berlin emotional data-base (EMODB) \cite{burkhardt2005database} and Ryerson audio-visual database (RAVDESS) \cite{livingstone2018ryerson} were used to evaluate speech emotion performance of our and baseline methods. Two key factors of both selected datasets are the difference in data size and language vocalization that can prove the performance of test methods. EMODB is a German speech in Berlin with 535 utterances and RAVDESS is English speech with 1440 utterances. EMODB dataset was recorded by male and female voices and contained seven different emotions: happiness, sadness, angry, neutral, fear, boredom, and disgust while RAVDESS dataset has one more emotion than EMODB, that is the calm emotion. Here, each dataset was divided into three subsets: 80\% for training set, 10\% for validation set, and 10\% for test set.

\subsubsection{Parameter settings for learning model}
All learning models were set up with the following parameter settings. Learning rate (LR) is very important in deep learning, when compared to step rate to find the minimum gradient. Generally, a high LR may make it over the minimum point. On the other hand, a low LR may take a long time to reach the goal. Here, we choose Adam optimizer for our experiments. Its learning rate and maximum epoch were set to 0.001 and 150, respectively. In addition, plateau strategy was used for reducing the LR and for avoiding overstepping the minimum point. In this case, we set the minimum LR to 0.00001. Batch size of models was set to 10. Finally, if an error value tends to increase, the early stopping criteria is active and then take the model weight with the previous minimum error.

\vspace{-10pt}
\begin{table}
\setlength\extrarowheight{2pt}
\caption{A performance comparison of DeepResLFLB and baseline methods with LMS feature, tested on Berlin EMODB dataset.
}\label{emodb_lms}
\begin{tabularx}{\textwidth}{|C|C|C|C|C|}
\hline
Method &  Accuracy & Precision & Recall & F1-score\\
\hline
Demircan \cite{demircan2018application} & 0.6755$\pm$0.0351 & 0.7549$\pm$0.0375 & 0.6295$\pm$0.0346 & 0.6295$\pm$0.0346 \\
1D-LFLB \cite{zhao2019speech} & 0.7577$\pm$0.0241 & 0.7609$\pm$0.0224 & 0.7574$\pm$0.0318 & 0.7514$\pm$0.0275 \\
2D-LFLB \cite{zhao2019speech} & 0.8269$\pm$0.0214 & 0.831$\pm$0.0228 & 0.824$\pm$0.0215 & 0.8233$\pm$0.0.0233 \\
{ DeepResLFLB} & {\bfseries 0.8404$\pm$0.0225} & {\bfseries 0.8481$\pm$0.0225} & {\bfseries 0.8298$\pm$0.0236} & {\bfseries 0.8328$\pm$0.0244} \\
\hline
\end{tabularx}
\end{table}
\vspace{-20pt}

\vspace{-10pt}
\begin{table}
\caption{A performance comparison of DeepResLFLB and baseline methods with LMSDDC feature, tested on Berlin EMODB dataset.
}\label{emodb_lmsddc}
\begin{tabularx}{\textwidth}{|C|C|C|C|C|}
\hline
Method &  Accuracy & Precision & Recall & F1-score\\
\hline
Demircan \cite{demircan2018application} & - & - & - & -  \\
1D-LFLB \cite{zhao2019speech} & 0.8355$\pm$0.0186 & 0.8385$\pm$0.0170 & 0.8313$\pm$0.0205 & 0.8322$\pm$0.0198 \\
2D-LFLB \cite{zhao2019speech} & 0.8754$\pm$0.0232 & 0.8802$\pm$0.0237 & 0.8733$\pm$0.0237 & 0.8745$\pm$0.0226 \\
{DeepResLFLB} & {\bfseries 0.8922$\pm$0.0251 } & {\bfseries 0.8961$\pm$0.0212 } & {\bfseries  0.8856$\pm$0.0322 } & {\bfseries 0.8875$\pm$0.0293 } \\
\hline
\end{tabularx}
\end{table}
\vspace{-30pt}

\begin{table}[h!]
\caption{A performance comparison of DeepResLFLB and baseline methods with LMS feature, tested on RAVDESS dataset.
}\label{ravdess_lms}
\begin{tabularx}{\textwidth}{|C|C|C|C|C|}
\hline
Method &  Accuracy & Precision & Recall & F1-score\\
\hline
Demircan \cite{demircan2018application} & 0.7528$\pm$0.0126 & 0.7809$\pm$0.0109 & 0.7422$\pm$0.0076 & 0.7479$\pm$0.0114 \\
1D-LFLB \cite{zhao2019speech} & 0.9487$\pm$0.0138 & 0.9491$\pm$0.0134 & 0.948$\pm$0.0123 & 0.9478$\pm$0.0133 \\
2D-LFLB \cite{zhao2019speech} & 0.9456$\pm$0.0128 & 0.9438$\pm$0.0136 & 0.946$\pm$0.0129 & 0.9442$\pm$0.0135 \\
{ DeepResLFLB} & {\bfseries 0.9602$\pm$0.0075 } & {\bfseries 0.9593$\pm$0.0072 } & {\bfseries 0.9583$\pm$0.0066 } & {\bfseries 0.9584$\pm$0.0071 } \\
\hline
\end{tabularx}
\end{table}

\begin{table}[h!]
\caption{A performance comparison of DeepResLFLB and baseline methods with LMSDDC feature, tested on RAVDESS dataset.
}\label{ravdess_lmsddc}
\begin{tabularx}{\textwidth}{|C|C|C|C|C|}
\hline
Method &  Accuracy & Precision & Recall & F1-score\\
\hline
Demircan \cite{demircan2018application} & - & - & - & - \\
1D-LFLB \cite{zhao2019speech} & 0.9367$\pm$0.0225 & 0.9363$\pm$0.0196 & 0.9352$\pm$0.0218 & 0.9347$\pm$0.0217 \\
2D-LFLB \cite{zhao2019speech} & 0.9466$\pm$0.0159 & 0.9475$\pm$0.0171 & 0.9441$\pm$0.0162 & 0.9449$\pm$0.0168 \\
{DeepResLFLB} & {\bfseries 0.949$\pm$0.0142 } & {\bfseries 0.9492$\pm$0.0143 } & {\bfseries 0.9486$\pm$0.016 } & {\bfseries 0.9484$\pm$0.0154 } \\
\hline
\end{tabularx}
\end{table}

\begin{table}[h!]
\centering
\caption{A comparison of a number of parameters in DeepResLFLB and LFLB models, tested on EMODB and RAVDESS datasets.
}\label{parameters}
\begin{tabularx}{\textwidth}{|C|C|C|C|C|}
\hline
Method & \multicolumn{2}{c|}{EMODB} & \multicolumn{2}{c|}{RAVDESS}\\
\cline{2-5}
 & LMS & LMSDDC & LMS & LMSDDC\\
\hline
2D-LFLB \cite{zhao2019speech} & 260544 & 262272 & 260544 & 262272 \\
{DeepResLFLB} & {\bfseries 156068} & {\bfseries 163268} & {\bfseries 156074} & {\bfseries 164608} \\
\hline
\end{tabularx}
\vspace{-15pt}
\end{table}

\subsubsection{Result discussion}
Based on dataset preparation and parameter setup for learning models, all experiments were used 5-fold validation. Tables 1 and 2 shows performance comparison between LMS and LMSDDC features, respectively, tested on EMODB dataset. It can be seen that LMSDDC feature (Table \ref{emodb_lmsddc}) provided the improvement of accuracy, precision, recall, and F1-score, when compared with LMS feature (Table \ref{emodb_lms}). In the same way, when the same learning models with different features, LMS and LMSDDC, were tested on RAVDESS dataset, as shown in Tables \ref{ravdess_lms} and \ref{ravdess_lmsddc}, the evaluation results were comparable, not much of an improvement. One of the main reasons is that RAVDESS has less speech variation than EMODB, as reported by Breitenstein research \cite{breitenstein2001contribution}. The less variation of speech leads to the lower quality of features. This made no difference in quality of LMS and LMSDDC features. These results have proved that the LMSDDC feature extracted with three components of human emotions: glottal flow, prosody, and human hearing, usually provided wider speech band frequencies, can improve the speech emotion recognition, especially with high speech variation datasets.

\begin{figure}
\includegraphics[width=\textwidth]{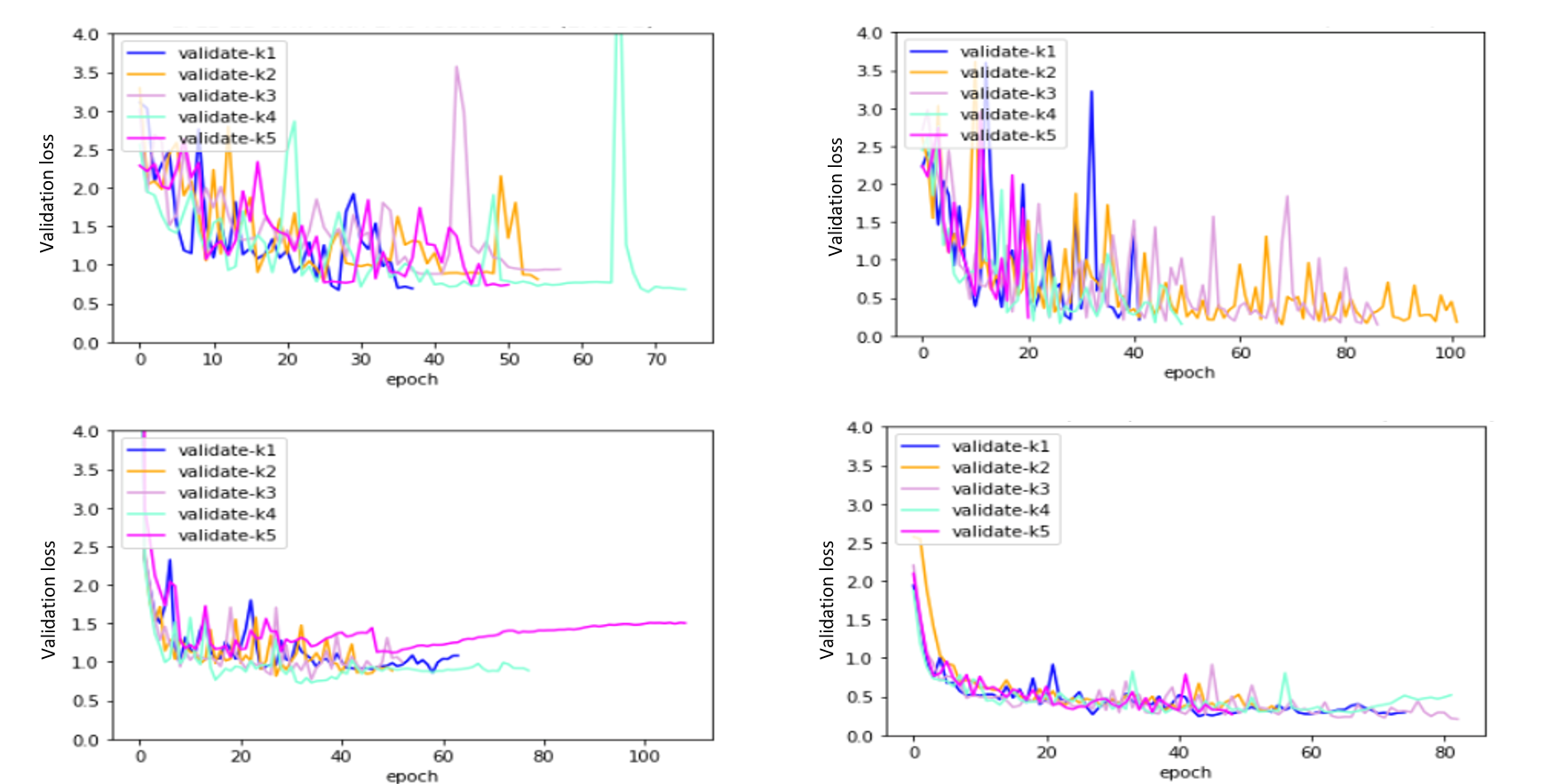}
\caption{Validation loss of learning models: conventional LFLB tested on EMODB (top-left), DeepResLFLB tested on EMODB (bottom-left), conventional LFLB tested on RAVDESS (top-right), and DeepResLFLB tested on RAVDESS (bottom-right). Note that only LMS feature was used to test in this experiment.} 
\label{loss}
\vspace{-23pt}
\end{figure}

When considering the efficiency of the learning model, Tables \ref{emodb_lms}, \ref{emodb_lmsddc}, \ref{ravdess_lms}, and \ref{ravdess_lmsddc} show that DeepResLFLB outperforms the baselines with the highest accuracy, precision, recall, and F1-score. This achievement proved that a learning sequence of DeepResLFLB, imitated from ``repeatedly reading" concept of human, is efficient.
In addition, DeepResLFLB can avoid a vanishing gradient problem and reduce resource-consuming. Fig.~\ref{loss} shows that DeepResLFLB had better validation loss and generalization; it can be seen from the graph that has less fluctuation than 2D-LFLB, and Table~\ref{parameters} reports that DeepResLFLB still used fewer parameters than the baseline model around 40\%. These results have proved that DeepResLFLB used residual deep learning by arranging its internal network as LFLB, BN, activation function, NAC, and deep layers can solve vanishing gradient and resource-consuming. Besides, when regarding resource-consuming between LMS and LMSDDC, LMSDDC parameters were slightly more than a baseline.

\vspace{-20pt}

\section{Conclusion}
This paper has described a DeepResLFLB model and LMSDDC feature for speech emotion recognition. The DeepResLFLB was redesigned from LFLB based on the `repeatedly reads' concept while the LMSDDC was emotional feature extracted from speech signals based on human glottal flow and human hearing. Performance of our model and emotional feature was tested on two well-known databases. The results show that the DeepResLFLB can perform better than baselines and use fewer resources in learning layers. In addition, the proposed LMSDDC can outperform conventional LMS.  

Although DeepResLFLB presented in this paper have provided better performance in speech emotion recognition, many aspects still can be improved, especially activation function. In future work, we will apply different kinds of activation function in each section of neural network; this will improve the performance of DeepResLFLB.

\vspace{-3pt}

\section*{Acknowledgments}
We would like to thank Science Research Foundation, Siam Commercial Bank, for partial financial support to this work.
%
%
%
\bibliographystyle{splncs04}
\bibliography{bibliography.bib}
\end{document}